\documentclass[twocolumn,tighten]{aastex61}
\usepackage{amsmath}

\shorttitle{Can the BZ mechanism power steady jets?}
\shortauthors{King \& Pringle}

\begin{document}

\title{Can the Blandford-Znajek mechanism power steady jets?}

\correspondingauthor{A.R. King}
\email{ark@leicester.ac.uk}

\author[0000-0002-2315-8228]{A.R. King}
\affiliation{School of Physics and Astronomy, University of Leicester, Leicester, LE1 7RH, UK}
\affiliation{Astronomical Institute Anton Pannekoek, University of Amsterdam, Science Park 904, NL-1098 XH Amsterdam, Netherlands}
\affiliation{Leiden Observatory, Leiden University, Niels Bohrweg 2, NL-2333 CA Leiden, Netherlands}

\author[0000-0002-2137-4146]{J.E. Pringle}
\affiliation{Institute of Astronomy, University of Cambridge, Madingley Road, Cambridge CB3 OHA, UK}

\begin{abstract}

We consider the Blandford-Znajek (BZ) mechanism for extracting black hole spin energy to drive astrophysical jets. Analyses of the BZ mechanism 
generally take no account of any electric charge on the black hole. 
But, as noted by Wald and others, if the medium surrounding the black hole is an ionised plasma with mobile charges, then a spinning hole quickly acquires an electric charge. The effect of this charge is to nullify the electric field structures which drive the BZ mechanism. Since jets are now observed in a wide variety of classes of accreting objects, most of which do not contain a central black hole, it seems likely that the jet driving mechanism in all astrophysical objects uses energy directly from the accretion disc, rather than black hole spin.

\end{abstract}

\keywords{astrophysical jets -- accretion discs -- black hole physics \vspace{0.5in}}

\section{Introduction}
\label{intro}

Early maps of double--lobe radio galaxies (e.g. Mitton \& Ryle, 1969) showed amorphous blobs of radio emission 
symmetrically placed each side of the central galaxy. Rees (1971) suggested
that an unknown object in the galactic nucleus channels energy to the radio lobes through jets. There is now almost universal agreement that the central object in radio galaxies 
is a supermassive black hole (Rees 1984), and that the high energy activity in the nucleus is powered by accretion (Salpeter, 1964), most likely through an accretion disc (Lynden--Bell, 1969). Thus the relativistic jets seen to emanate from the nucleus are ultimately powered by accretion of matter on to the central black hole (see the reviews by Begelman, Blandford \& Rees, 1984; Heckman \& Best, 2014; Blandford, Meier \& Readhead, 2019).

Blandford \& Znajek (1977, hereafter BZ77) proposed a radical new mechanism in which the jets  from galactic nuclei are powered by direct electromagnetic extraction of the spin energy 
from the central black hole. This is the Blandford--Znajek (BZ) mechanism. 

In Section~\ref{sec:BZbasic} we provide a brief overview of this mechanism, which invokes a spinning black hole situated in an aligned magnetic field. We note
 that the central black hole is tacitly assumed always to have negligible electric charge,
that is, any current entering the hole is assumed to be balanced by a precisely opposing current entering elsewhere, so that there is an effective current through the hole.
In Section~\ref{sec:Addcharge} we draw attention to the analysis by Wald (1974; see also King et al., 1975; Petterson, 1975; Gibbons et al., 2013) in which he finds that if the black hole is surrounded by a plasma which contains mobile charges, it will acquire a specific electric charge. The charge is exactly such as to render the BZ mechanism inoperative. We provide a brief discussion in Section~\ref{Discuss}.

\section{The Blandford--Znajek mechanism
\label{sec:BZ}}

\subsection{The basic mechanism; the uncharged black hole \label{sec:BZbasic}}

Black holes are generally assumed to have zero net electric charge. The reason is that if for example
a Schwarzschild (non--rotating) black hole is given a charge, the neighbourhood of the hole then acquires an electric field. If (and only if) charge separation is allowed, charged particles in the surrounding astrophysical plasma move parallel or antiparallel to the electric field. In this way the black hole selectively acquires charges so that it moves quickly towards zero net charge. Charge separation is well known to occur in electrical media in which the charge carriers are able to move independently. It occurs for example in electrolytes (Debye \& H\"uckel, 1923) and in ionized astrophysical plasmas (Salpeter, 1954), and it affects the rates of nuclear burning in stars
(e.g. Clayton, 1968). The timescale to reach zero net charge is typically quite short in realistic astrophysical environments (see, for example, the discussion by Cardoso et al., 2016). Thus, Blandford (1987) asserts that ``charged, Kerr--Newman black holes are irrelevant to astronomy''.

Soon after black holes were recognised as a realistic astrophysical possibility,
Wald (1974) devised an elegant method to 
compute the effect on the electromagnetic field structure when a black hole, with zero net charge,  is placed in a uniform magnetic field. King et al. (1975) extended this result to consider  more general aligned magnetic field structures. Wald (1974) showed that for a spinning (Kerr) black hole, with zero net electric charge, and with spin aligned with an external uniform magnetic field ${\bf B}$ in a vacuum, the 
spin induces an electric field with ${\bf E \cdot B \neq 0}$ (for a classical analogy, see Ruffini \& Treves, 1973).
Wald assumed a field uniform at infinity, but King et al. (1975) showed that the field structure near the hole is essentially 
the same for any realistic aligned field. Wald also noted that an electric field with $\bf E \cdot B \neq 0$ would lead to movement of any external charged particles. King et al. (1975) showed that the induced electric field structure is such that $\bf E \cdot B$  has opposite signs near the poles and near the equator of the spinning hole. This implies that  charges of one sign would be attracted towards the poles, whereas charges of the opposite sign would be attracted to a band near the equator.
  
This result led BZ77 to consider a spinning black hole at the centre of an accretion disc with the disc providing the required currents to give rise to an aligned magnetic field.
The Bardeen--Petterson effect (Bardeen \& Petterson, 1975) ensures that the central disc and black hole spin are generally aligned, and so by symmetry the magnetic field should also be aligned with the spin. 
BZ77 then noted that if the black hole is surrounded by a conducting medium the structure of the induced electric field found by Wald (1974) and by King et al. (1975)  must produce an effective electric current through the black hole, returning through the surrounding medium (see, for example, Thorne \& Blandford, 1982; Blandford, 1987). Any dissipation of this current's energy in the surrounding material then taps the spin energy of the black hole.

This is the BZ--mechanism. It provides, in principle, a mechanism for the continuous extraction of energy from a spinning black hole. BZ77 speculated that it could be used to power the astrophysical jets seen to emanate from active galactic nuclei (see also Blandford, Meier \& Readhead, 2019). The luminosity $L_{\rm BZ}$ 
 produced by this process is, on dimensional grounds, 
\begin{equation}
\label{BZpower}
L_{\rm BZ} \sim a^2 \times \frac{B^2}{8 \pi} \times \frac{ 4 \pi R_{\rm s}^3}{3} \times \frac{c}{R_s}.
\end{equation}
for a black hole of mass $M$, Schwarzschild radius $R_{\rm s} \sim 2GM/c^2$, dimensionless spin parameter 
$a$, where $0 \le a^2 \le 1$, placed in a magnetic field of strength  $B$, i.e. approximately the magnetic energy contained by the formal `volume' of the hole, emitted every light crossing time of the hole, and moderated by the amount of spin energy available.

For a `fiducial' field strength (Begelman, Blandford \& Rees, 1984; but see Ghosh \& Abramowicz, 1997) of
\begin{equation}
B \sim 2 \times 10^4  M_8^{-1/2} \: {\rm G},
\end{equation}
this gives a luminosity 
\begin{equation}
L_{\rm BZ} \sim 2 \times 10^{45} a^2 M_8 \: {\rm erg\, s^{-1}},
\label{BZpower2}
\end{equation}
where $M_8$ is the mass of the black hole in units of $10^8 M_\odot$. This is comparable to the Eddington luminosity
\begin{equation}
L_E \sim 1.3 \times 10^{46} M_8 \: {\rm erg\, s^{-1}}.
\end{equation}

BZ77 investigated this process using the force--free approximation in a charge--separated plasma in which particle inertia and interparticle collision terms can be ignored (see also Komissarov, 2004). Since then a number of authors have investigated this mechanism assuming that the surrounding medium could be modelled using the MHD approximation in which collision terms dominate, often by numerical means (see for example the reviews by Davis \& Tchekhovskoy, 2020; Komissarov \& Porth, 2021). However, in all these investigations, it is implicitly assumed that the movements of charges in any surrounding plasma do not permit a change in the net electrical charge of the black hole.

\medskip

\subsection{The acquisition of charge and its implication
\label{sec:Addcharge}}

The discussion of Section~\ref{sec:BZbasic} above assumes, as is usually the case in astronomy, that the spinning black hole has negligible net electric charge. 
However, in his seminal paper, Wald (1974) reasoned that a black hole would be surrounded by a standard astrophysical plasma in which the possibility of charge separation would exist. 
In this case, it is to be expected that the electric field drives the charges in such a way as to lead to a drop in electric potential along the magnetic field lines (cf. Komissarov, 2004).
He argued further that for a spinning black hole in an aligned magnetic field ${\bf B}$ the movements of charge carriers (i.e. currents)  induced by the electric field with ${\bf E \cdot B \neq 0}$, together with the fact that the charge carriers are individually mobile, would lead to the hole selectively accreting net charge in such a way as to nullify the effects of the electric field. 

Specifically, Wald showed that the net charge reaches the value
\begin{equation}
Q = 2BJ 
\label{charge}
\end{equation}
in geometrized units (where $J = Ma$ is its total angular momentum). 
Then the charge on the hole remains constant, removing the need for currents, which might then drive the BZ effect.


This result was confirmed and generalised by Petterson (1975), who showed that the precise
value $Q'$ of the critical charge in units of $Q$ depends on the distribution of the source
currents of the magnetic field.

The charge $Q'$ is utterly negligible ($Q'^2 \ll M^2$) in gravitational terms (Wald, 1974; 
Zajacek et al, 2018; Zajacek \& Tursunov, 2019), so the spacetime metric is still to a high approximation uncharged Kerr, in agreement with the remark by Blandford (1987) quoted above. In particular the motion of uncharged particles is effectively identical to the case $Q'=0$. But charged particle motion is very different, and strongly influenced by the charge $Q'$. This is another illustration of how extremely weak gravity is by comparison with electromagnetism.
Thus, if the surrounding conducting medium is treated as a realistic space plasma which permits a net flow of charge into the black hole, then rather than providing a continuous process for removing spin energy from the hole, the induced electric currents are an initial transient effect which continues only until the black hole acquires the charge $Q' \simeq 2BJ$ (cf Zajacek et al.,  2018; Zajacek \&  Tursunov 2019)\footnote{For all conceivable boundary conditions far from the hole its net charge tends monotonically to $Q'$.}.

The energy released in this transient is, using the above numbers, 
\begin{equation}
E_Q \sim L_{\rm BZ} \times \frac{R_{\rm s}}{c} \sim 2 \times 10^{48} a^2 M_8^2 \: {\rm erg},
\end{equation}
i.e. the transient emits the luminosity $L_{\rm BZ}$ for a time $\sim R_{\rm s}/c \sim 10^3M_8\,{\rm s}$.
Once the black hole acquires this charge, the torque on it vanishes and no more spin energy can be extracted. 

More recently numerical modelling of the BZ--mechanism has been
undertaken using Particle--In--Cell (PIC) plasma methods, which permit
independent mobility of individual charges. In principle these
techniques should be able to test Wald’s fundamental hypothesis that a
spinning black hole immersed in a magnetic field should acquire a net
charge. The same technique has also been applied to ionised plasma
surrounding rotating neutron stars (e.g. Kalapotharakos et al., 2018).
However, in contrast to the neutron star case, where Kalapotharakos et
al. (2018) treat the inner boundary with some care, noting that they
ensure “current closure of charge carriers that reach the stellar
surface”, modelling of the black hole case is often done with inner
boundary conditions which either prevent or ignore the acquisition of
charge by the black hole. For example, Parfrey et al (2019, see also Crinquand et al., 2019) do not comment on charge acquisition, and
Hirotani et al. (2021) impose that
${\bf E\cdot B} = 0$ and that both the radial component of the electric field and the
meridional component of the magnetic field vanish at the inner 
boundary\footnote{Note added in proof: Parfrey (private communication) informs us that, during the timespan of the simulations presented in Parfrey et al (2019), the black hole both acquires an electric charge and exhibits an outflow of electromagnetic energy.}.

\section{Discussion}
\label{Discuss}

The BZ mechanism is a standard cited mechanism for producing steady jets in objects that contain accreting black holes -- galactic nuclei and some X--ray binaries. And indeed the process is often cited in papers which concern numerical MHD simulations of jets and outflows produced by magnetic accretion discs (see the review by Davis \& Tchekhovskoy, 2020). We have argued above, in line with the original ideas of Wald (1974; see also Gibbons et al., 2013) 
that in a realistic space plasma which permits a net flow of charge into the black hole
the BZ mechanism cannot tap the spin energy of a black hole continuously, and is therefore not a viable mechanism for powering continuous astrophysical jets.

This is primarily because the conducting medium surrounding the black hole should be treated as an astrophysical plasma with mobile charge carriers. When charge separation is allowed, along with the
possibility of a net flux of charge into the black hole,
any spinning black hole quickly acquires the net electrical charge $Q'$, and the electric fields which drive the currents required for the BZ mechanism are nullified. The same process of charge separation which ensures that a non--rotating black hole has zero charge also ensures that a rotating hole acquires the charge $Q'$ that makes the BZ mechanism inoperative.
We have noted that these ideas need to be tested, for example using PIC
plasma simulation techniques. For example, it might be that collective plasma effects serve to
counteract the tendency of the black hole to acquire charge\footnote{We
thank the referee for stressing this possibility.}.

There are, of course, many other kinds of astrophysical objects which do not contain black holes and nevertheless produce jets (Burgarella et al., 1993; Smith, 2012). The jets emitted by young stellar objects are particularly spectacular (see the review by Ray \& Ferreira, 2021). Thus the application of Occam's 
razor\footnote{`Do not multiply hypotheses', or put simply, `don't invent two theories for the same thing'.}
has long suggested that the BZ mechanism, even if it were viable, is in fact not required for the production of astrophysical jets (Livio, 1997; see also Pringle, 1993; Price, Pringle \& King, 2003). In addition, Russell, Gallo \& Fender (2013) have shown that the jet production mechanism in binary X--ray sources is not consistent with the prediction of the BZ mechanism that the jet power should depend on the the square of the dimensionless jet spin parameter (see equation~\ref{BZpower2}). 

Thus, following Occam, if one is forced to choose a single mechanism capable of producing all astrophysical jets which emanate from accreting objects, then the most likely choice would be some form of MHD process resulting from poloidal magnetic field threading the accretion disc (Livio, 1997; Livio et al., 1999). A mechanism like this is already discussed by Blandford \& Znajek (1977), and early ideas on this process are given by Blandford \& Payne (1982)  and, in the protostellar case, by Pudritz \& Norman (1983, 1986).

\section*{Acknowledgments}

We thank Bob Carswell, Gary Gibbons, Chris Nixon, Colin Norman, Kyle Parfrey, Roger Blandford
and Roman Znajek for helpful comments,
and the referee for a thoughtful report.

\bigskip
\bigskip
\section*{ REFERENCES}
\bigskip
\noindent
Bardeen, J.A. \& Petterson, J.A., 1975, ApJL 195, L65

\noindent
Begelman, M.C., Blandford, R.D., \& Rees, M.J., 1984, Rev. Mod. Phys., 56, 255

\noindent
Blandford, R.D., 1987, in {\it Three Hundred Years of Gravitation}, Cambridge University Press, 
eds S. Hawking \& W. Israel,  pp. 277 -- 329

\noindent
Blandford, R.D.,  Meier, D.  \& Readhead, A.  2019, ARA\&A, 57, 467

\noindent
Blandford, R.D., \& Payne, D.G., 1982, MNRAS, 199, 883

\noindent
Blandford, R.D., \&  Znajek, R. L. 1977, MNRAS 179, 433

\noindent
Burgarella, D., Livio, M., \& O'Dea, C.P. (eds), 1993, {\it Astrophysical Jets}, Cambridge University Press

\noindent
Cardoso, V., Macedo C. F. B., Pani, P., \& Ferrari, V., 2016, JCAP, 054

\noindent
Clayton, D.D., 1968, {\it Principles of Stellar Evolution and Nucleosynthesis}, Univ. of Chicago Press

\noindent
Crinquand, B., Cerutti, B., Dubus, G., Parfrey, K., Philippov, A.,  2021, A\&A, 650, A163

\noindent
Debye, P., \&  H\"uckel, E., 1923, Physikalische Zeitschrift, 24, 185

\noindent
Ghosh, P, \& Abramowicz, M.A., 1997, MNRAS, 292, 887

\noindent
Gibbons, G.~W., Mujtaba, A.~H., \& Pope, C.~N.\ 2013, Classical and Quantum Gravity, 30, 125008

\noindent
Heckman, T.M., \& Best, P. N., 2014, ARA\&A, 52, 589

\noindent
Hirotani, K., Krasnopolsky, R., Shang, H., Nishikawa, K., Watson, M.,  2021, ApJ, 908, 88

\noindent
Kalapotharakos, C., Brambilla, G. Timokhin, A. Harding, A. K. \& Kazanas, D.,
2018, ApJ, 857, 44

\noindent
King, A.R., Lasota, J.P., \& Kundt, W., 1975, Phys. Rev. D,12, 3037

\noindent
Komissarov, S.S., 2004, MNRAS, 350, 427
 
\noindent
Komissarov, S.S., \& Porth, O., 2021, NewAR, 92, 101610

\noindent
Livio, M.,  1997,  in {\it Accretion Phenomena and Related Outflows} eds D.T. Wickramasinghe,
L. Ferrario, \& G.V. Bicknell, ASP Conference Series 121, 845

\noindent
Livio., M., Ogilvie, G,I., \& Pringle, J.E., 1999, ApJ, 512, 100

\noindent
Lynden--Bell, D., 1969, Nature, 223, 690L

\noindent
Mitton., S., \& Ryle, M., 1969, MNRAS, 146, 221

\noindent
Parfrey, K., Philippov, A., Cerutti, B., 2019, Phys. Rev. Lett. 122c5101

\noindent
Petterson, J.~A.\ 1975, Phys Rev D,12, 2218

\noindent
Price, D.J., Pringle, J.E., \& King, A.R., 2003, MNRAS 339, 1223

\noindent
Pringle, J.E., 1993,  in {\it Astrophysical Jets}, Cambridge University Press, eds Burgarella, D., Livio, M., \& O'Dea, C.P. 

\noindent
Pudritz, R.E., \& Norman, C.A., 1983, ApJ, 274, 677

\noindent
Pudritz, R.E., \& Norman, C.A., 1986, ApJ, 301, 571\noindent

\noindent
Ray, T.P, \& Ferreira, J.  2021, NewAR, 93, 101615

\noindent
Rees, M.J., 1971, Nature, 229, 312

\noindent
Rees, M.J., 1984, ARA\&A, 22, 471

\noindent
Ruffini, R. \& Treves, A.\ 1973, \aplett, 13, 109

\noindent
Russell, D.M., Gallo, E., \& Fender, R.P., 2013, MNRAS, 431, 405

\noindent
Salpeter, E.E., 1954, Aus J. Phys., 7, 373

\noindent
Salpeter, E.E., 1964, ApJ, 140, 796

\noindent
Smith, M.D.,  2012, {\it Astrophysical Jets and Beams} Cambridge University Press

\noindent
Thorne, K.S., \& Blandford, R.D., 1982, in {\it Proc IAU Symp 97} eds. Heeschen, D.S. \& Wade, C.M., 255

\noindent
Wald, R.M., 1974, Phys Rev D, 10, 1680

\noindent
Zajacek, M., Tursunov, A., Eckhart, A., Britze, S., 2018, MNRAS, 480, 4408

\noindent
Zajacek, M., \& Tursunov, A., 2019, The Observatory, 139, 231

\end{document}